\documentclass[journal]{IEEEtran} 
\IEEEoverridecommandlockouts
\usepackage[left=0.625in,right=0.625in,top=0.7in,bottom=1.1in]{geometry}
\usepackage[dvipsnames]{xcolor}
\usepackage{cite}
\usepackage{amsmath,amssymb,amsfonts}
\usepackage{textcomp}
\usepackage{xcolor}
\usepackage{float}
\usepackage{cite}
\usepackage{graphicx}
\usepackage{placeins}
\usepackage{textcomp}
\usepackage{multirow}
\usepackage{xcolor}
\usepackage{wrapfig}
\usepackage{algpseudocode}
\usepackage{algorithm}
\usepackage{caption}
\usepackage{subcaption}
\usepackage{comment}

\usepackage[noabbrev]{cleveref}

\usepackage{tikz}
\usepackage{import}
\usepackage{pgfplots}
\usetikzlibrary{calc}
\usetikzlibrary{spy}
\usetikzlibrary{matrix}
\usetikzlibrary{backgrounds}
\pgfplotsset{compat = newest}

\usetikzlibrary{shapes.misc, positioning}
\usetikzlibrary{shadows,arrows,shapes,calc,positioning,decorations.pathreplacing,fit}
\def\BibTeX{{\rm B\kern-.05em{\sc i\kern-.025em b}\kern-.08em
		T\kern-.1667em\lower.7ex\hbox{E}\kern-.125emX}}

\DeclareMathOperator*{\argmax}{\mathrm{argmax}}

\makeatletter
\def\endthebibliography{%
  \def\@noitemerr{\@latex@warning{Empty `thebibliography' environment}}%
  \endlist
}
\makeatother

\begin{document}

\pgfplotsset{
    standard/.style={
    axis line style = thick,
    grid = both,
    }
}

\pagestyle{empty}

\title{On the Robustness of Deep Learning-aided Symbol Detectors to Varying Conditions and Imperfect Channel Knowledge\\

}

\author{
    \IEEEauthorblockN{ Chin-Hung Chen$^{\star}$}
    , \IEEEauthorblockN{Boris Karanov$^{\star}$}%
    , \IEEEauthorblockN{Wim van Houtum$^{\star\dagger}$}%
    , \IEEEauthorblockN{Wu Yan$^{\dagger}$}%
    , \IEEEauthorblockN{Alex Young$^{\dagger}$}%
    , \IEEEauthorblockN{Alex Alvarado$^{\star}$}%
    \\
    \IEEEauthorblockA{\textit{$^{\star}$Eindhoven University of Technology, Eindhoven 5600 MB, The Netherlands}}\\
    \IEEEauthorblockA{\textit{$^{\dagger}$NXP Semiconductors, High Tech Campus 46, Eindhoven 5656 AE, The Netherlands}}\\
    \IEEEauthorblockA{c.h.chen@tue.nl}
}

\maketitle
\thispagestyle{empty}

\begin{abstract}
Recently, a data-driven Bahl-Cocke-Jelinek-Raviv (BCJR) algorithm tailored to channels with intersymbol interference has been introduced. This so-called BCJRNet algorithm utilizes neural networks to calculate channel likelihoods. \mbox{BCJRNet} has demonstrated resilience against inaccurate channel tap estimations when applied to a time-invariant channel with ideal exponential decay profiles. However, its generalization capabilities for practically-relevant time-varying channels, where the receiver can only access incorrect channel parameters, remain largely unexplored. The primary contribution of this paper is to expand upon the results from existing literature to encompass a variety of imperfect channel knowledge cases that appear in real-world transmissions. Our findings demonstrate that \mbox{BCJRNet} significantly outperforms the conventional BCJR algorithm for stationary transmission scenarios when learning from noisy channel data and with imperfect channel decay profiles. However, this advantage is shown to diminish when the operating channel is also rapidly time-varying. Our results also show the importance of memory assumptions for conventional BCJR and BCJRNet. An underestimation of the memory largely degrades the performance of both BCJR and BCJRNet, especially in a slow-decaying channel. To mimic a situation closer to a practical scenario, we also combined channel tap uncertainty with imperfect channel memory knowledge. Somewhat surprisingly, our results revealed improved performance when employing the conventional BCJR with an underestimated memory assumption. BCJRNet, on the other hand, showed a consistent performance improvement as the level of accurate memory knowledge increased.

\end{abstract}

\begin{IEEEkeywords}
    BCJR algorithm, intersymbol interference, machine learning, neural networks, symbol detection
\end{IEEEkeywords}

\section{Introduction} \label{sec:intro}
The utilization of machine learning algorithms and neural networks (NN) in communication systems has drawn significant attention, primarily due to their remarkable capacity for universal function approximation~\cite{Simeone2018,OShea2017}. For instance, end-to-end learning systems, tailored for data transmission over channels for which the optimal transmitter-receiver pair is unknown, have been proposed for both wireless and fiber-optic communications~\cite{Dörner18,Karanov18,Karanov19}. These systems often rely on deep multi-layer NNs \cite{Aoudia2018,Hao2018,OShea2019,Karanov2020}. However, a notable limitation of deploying these NNs in practical applications is their need for continuous optimization, which is often a complex task.

The complexity and optimization effort can be greatly reduced by confining machine learning only to the receiver side. For example, the authors of~\cite{Farsad2018} introduced a fully deep learning-based sequence detector using recurrent NNs. Also, a low-complexity solution that integrates NNs into receiver design based on message-passing algorithms has been shown to require significantly less training data~\cite{VitNet,Nachmani18,Shlezinger21,Shlezinger22,Shlezinger23}. Such an approach preserves the conventional receiver structure, such as the Bahl-Cocke-Jelinek-Raviv (BCJR)~\cite{bcjr74} or the Viterbi~\cite{Forney} algorithm, with the added capability to learn channel likelihoods directly from data. This approach alleviates the need for perfect knowledge of the underlying communication channel, often unavailable in real-world systems.

In a recent study, a machine learning-aided implementation of the maximum a-posteriori probability (MAP) detector, called BCJRNet, was introduced~\cite{Shlezinger21}. BCJRNet employs a relatively small feedforward NN and a Gaussian mixture model (GMM)-based density estimator~\cite{mclachlan200001} to compute the likelihoods required for the BCJR algorithm. The conventional BCJR detector requires perfect channel state information (CSI) knowledge to achieve optimal performance. In contrast, \mbox{BCJRNet} uses the aforementioned NN and GMM frameworks to learn the likelihoods from labeled training data. The authors of \cite{Shlezinger21} showed that when the training data originates from a channel that is identical to the one used in the detection (testing) stage, BCJRNet's performance closely aligns with the optimal conventional BCJR detector, which assumes perfect CSI. Notably, \cite{Shlezinger21} showed that the data-driven detector significantly outperforms the conventional implementation based on inaccurate CSI estimation when operating on an intersymbol interference (ISI) channel where the taps follow an ideal exponential decay profile. This result highlights the great potential of such machine learning-aided systems in adapting the established BCJR algorithm to CSI uncertainties. 

The available results in the literature~\cite{Shlezinger21,Shlezinger22,Shlezinger23} mainly consider the generalization capabilities of different data-driven detectors only when operated with a stationary ISI model. However, in real-world applications, the channel used for transmission is usually time-varying. The channel is also likely to deviate in different ways from the estimated/learning channel \cite{Chagnon18,Karanov2020,Karanov2021} (i.e., imperfect CSI). Thus, a comprehensive examination of the robustness and versatility of these detectors is essential.

The main contribution of this paper is in extending the investigation of BCJRNet in~\cite{Shlezinger21} to a broader set of use-case scenarios. We concentrate on the performance and versatility of BCJRNet and the conventional BCJR detector operated in rapidly time-varying conditions with different degrees of CSI knowledge. More specifically, the following different scenarios are considered: imperfect knowledge of the channel memory, deviation in the exponential behavior of the channel response, error in the channel tap estimation, as well as operation over rapidly time-varying channels with and without perfect knowledge of the channel memory. Generally, BCJRNet exhibits superior generalization capabilities compared to conventional BCJR. However, this advantage diminishes notably when time-varying channels are considered. Additionally,  memory assumptions are shown to significantly influence the system performance of both detectors.

The paper is structured as follows: Sec.~\ref{sec:sys_model} introduces the system model for communication over ISI AWGN channels and the BCJR algorithm for MAP detection. Sec.~\ref{sec:exp_design} details the simulation setup and the operation and learning scenarios for the considered system. Sec.~\ref{sec:results} presents numerical results and corresponding comparisons. Sec.~\ref{sec:conc} summarize the paper with concluding remarks.

\section{System Model and Receiver Design} \label{sec:sys_model}
In this section, we briefly review the communication model with finite memory and the optimal conventional BCJR detector under this channel assumption. Then, we introduce the data-driven BCJRNet detector. This detector exploits NN to learn the CSI-related likelihood function, attaining a model-agnostic capability.
\subsection{Finite memory communication channel}
In this study, similar to~\cite{Shlezinger21,Shlezinger22,Shlezinger23}, we focus on a finite memory AWGN channel model with an input-output relationship given by
\begin{equation}\label{eq:channel_mod}
    Y_t = \sum_{l=1}^L h_{l,t} X_{t-l+1} + W_t,
\end{equation}
where $X_t\in\mathcal{X}$ represents the symbol transmitted at time index $t\in\{1,2,\ldots T\}$, where $\mathcal{X}$ is the constellation used for transmission and $T$ is the frame length. The channel output at time $t$ is denoted by $Y_t$ with the channel response denoted by $\{h_{1,t},...,h_{L,t}\}$ where $L$ is the channel memory. We assume the noise $W_t \in \mathcal{N}(0,\sigma_w^2)$ is independent of the transmitted symbols and follows a Gaussian distribution with zero mean and variance $\sigma_w^2$.

The conditional probability density function of the channel output sequence $\mathbf{Y}$ given its input sequence $\mathbf{X}$ for a finite-memory channel can be factorized as
\begin{equation} \label{eq:ISI_channel_factorization}
    p(\mathbf{y}|\mathbf{x}) = \prod_{t=1}^{T}p(y_t|\{x_l\}_{l=t-L+1}^t),
\end{equation}
where $y_t$ and $x_t$ are realizations of the received $Y_t$ and transmitted $X_t$ random variables, respectively. Note that $x_t=0$ for $t<0$.

For channels like the one in \eqref{eq:channel_mod}, it is common to define the channel state as $\mathbf{S}_t\triangleq\left\{X_{t},\ldots,X_{t-L+1}\right\} \in \mathcal{X}^L$ with the state transition probability 
\begin{equation} \label{eq:state_tran}
    p(\mathbf{s}_t|\mathbf{s}_{t-1}) = \begin{cases}
    p(x_t) = \frac{1}{M}, & (\mathbf{s}_t, \mathbf{s}_{t-1}) \in \mathcal{S} \\
    0, & \text{otherwise}.
\end{cases}
\end{equation}
Here, $\mathbf{s}_t$ and $\mathbf{s}_{t-1}$ are realizations of $\mathbf{S}_t$ and $\mathbf{S}_{t-1}$, respectively. $\mathcal{S}$ is the set of the pairs $(\mathbf{s}_t,\mathbf{s}_{t-1})$ corresponding to all states transitions driven by $x_t$. $M=|\mathcal{X}|$ denotes the cardinality of the constellation. We can now write the likelihood in the right-hand side of \eqref{eq:ISI_channel_factorization} as
\begin{equation}\label{eq:p(y|s)}
    p(y_t|\mathbf{s}_t=\{x_{t},\ldots,x_{t-L+1}\}) = \frac{1}{\sqrt{2\pi\sigma_w^2}}\exp{\frac{-(y_t-\mu_t)^2}{2\sigma_w^2}},
\end{equation}
where the accurate computation of the mean $\mu_t$ requires CSI knowledge in terms of $L$, and $\{h_{1,t},...,h_{L,t}\}$ since
\begin{equation}\label{eq:smean}
    \mu_t = \sum_{l=1}^L h_{l,t} x_{t-l+1}. 
\end{equation}


\subsection{MAP detection and BCJR algorithm}
The symbol detection strategy based on the MAP rule, which minimizes the symbol error rate, can be expressed as
\begin{equation} \label{eq:MAP_detection}
    \hat{x}_t = \argmax_{x_t\in \mathcal{X}}p(x_t|\mathbf{y}) = \argmax_{x_t\in \mathcal{X}}p(x_t,\mathbf{y}),
\end{equation}
with
\begin{equation}\label{eq:MAP_detection_2}
	p(x_t,\mathbf{y}) = \sum_{(\mathbf{s}_t, \mathbf{s}_{t-1}) \in \mathcal{S}} p(\mathbf{y},\mathbf{s}_t,\mathbf{s}_{t-1}).
\end{equation}

The BCJR algorithm efficiently computes the MAP rule in \eqref{eq:MAP_detection_2} using a factorization of the joint distribution for channels with finite memory. First, the function
\begin{equation} \label{eq:MAP_detection_gamma}
\begin{split}
    &f(y_t,\mathbf{s}_t,\mathbf{s}_{t-1}) \triangleq p(y_t|\mathbf{s}_t)p(\mathbf{s}_t|\mathbf{s}_{t-1}) \\
    &=\begin{cases}
		\frac{1}{M}p(y_t|\mathbf{s}_t), & (\mathbf{s}_t, \mathbf{s}_{t-1}) \in \mathcal{S}\\
		0, & \text{otherwise}
      \end{cases}
\end{split}
\end{equation}
is defined over the received symbol $y_t$ at time $t$ and the current and previous channel states $\mathbf{s}_t$ and $\mathbf{s}_{t-1}$. Then, we can compute the joint probability in the right-hand side of \eqref{eq:MAP_detection_2} by recursive message passing along a factor graph representation. In particular, we have
\begin{equation}\label{eq:factor_graph_message_passing}
    \begin{split}
    p(\mathbf{y},\mathbf{s}_t,\mathbf{s}_{t-1}) = \overrightarrow{\mu}(\mathbf{s}_{t-1})f(y_t,\mathbf{s}_{t},\mathbf{s}_{t-1})\overleftarrow{\mu}(\mathbf{s}_{t}),     
    \end{split}
\end{equation}
where the forward and backward messages are denoted by $\overrightarrow{\mu}$ and $\overleftarrow{\mu}$, respectively. These are recursively computed as
\begin{equation}\label{eq:forward_messages}
\begin{split}
    \overrightarrow{\mu}(\mathbf{s}_{i})=&\sum_{\mathbf{s}_{i-1} \in \mathcal{X}^L }f(y_i,\mathbf{s}_{i},\mathbf{s}_{i-1})\overrightarrow{\mu}(\mathbf{s}_{i-1}), \\
    &\quad \text{for } i = 1,2, \ldots, t     
\end{split}
\end{equation}
and  
\begin{equation}\label{eq:backward_messages}
\begin{split}
    \overleftarrow{\mu}(\mathbf{s}_{i}) =& \sum_{\mathbf{s}_{i+1}  \in \mathcal{X}^L}f(y_{i+1},\mathbf{s}_{i+1},\mathbf{s}_{i})\overleftarrow{\mu}(\mathbf{s}_{i+1}), \\
    &\quad \text{for } i = T-1,T-2, \ldots, t+1.    
\end{split}
\end{equation}
Combining the message passing computations from \eqref{eq:factor_graph_message_passing} with the MAP symbol detection rule given by \eqref{eq:MAP_detection} and \eqref{eq:MAP_detection_2}, we end up with the expression for the MAP estimate of the transmitted symbol given by
\begin{equation} \label{eq:MAP_estimate}
	\hat{x}_t = \argmax_{x_t\in \mathcal{X}}\sum_{(\mathbf{s}_t, \mathbf{s}_{t-1}) \in \mathcal{S}}\overrightarrow{\mu}(\mathbf{s}_{t-1})f(y_t,\mathbf{s}_{t},\mathbf{s}_{t-1})\overleftarrow{\mu}(\mathbf{s}_{t}).
\end{equation}

\subsection{Data-driven BCJR algorithm} \label{sec:data_driven_bcjr}
Unlike conventional BCJR, which requires full CSI knowledge to calculate the likelihoods by \eqref{eq:p(y|s)}. BCJRNet, as first proposed in~\cite{Shlezinger21}, uses training data and machine learning to compute the likelihoods $p(y_t|\mathbf{s}_t)$ for $t=1,2,\dots, T$. After this computation, the BCJR algorithm can be implemented by \eqref{eq:MAP_detection_gamma}--\eqref{eq:MAP_estimate}. In particular, as suggested in~\cite{Shlezinger21}, an NN is used to classify channel states from a received symbol. The optimized NN thus approximates the posterior probability $p^{\theta}(\mathbf{s}_t|y_t)$ ($\theta$ denotes the set of NN parameters). Then the desired likelihoods can be computed as
\begin{equation} \label{eq:Bayes_rule_ML}
	p^{\theta,\phi}(y_t|\mathbf{s}_t) = \frac{p^{\theta}(\mathbf{s}_t|y_t) p^{\phi}(y_t)}{p(\mathbf{s}_t)}=\frac{p^{\theta}(\mathbf{s}_t|y_t) p^{\phi}(y_t)}{{1}{/M^L}}.
\end{equation}
The network is optimized using pairs of received symbols and corresponding (labeled) channel inputs, i.e., working with the dataset $\{y_t,x_t\}_{t=1}^{T_{\text{data}}}$. Using a softmax NN output, the network parameters $\theta$ are optimized by minimizing:
\begin{equation} \label{eq:nn_para}
	\mathcal{L}(\theta) = \frac{1}{T_{\text{data}}}\sum_{t=1}^{T_{\text{data}}}-\log{p^{\theta}(\left[x_{t},\ldots,x_{t-L+1}\right]|y_t)}.
\end{equation}
The marginal probability $p^\phi(y_t)$ required in \eqref{eq:Bayes_rule_ML} is approximated via GMM as in~\cite{Shlezinger21}.

		

\section{Simulation Setup} \label{sec:exp_design}

\begin{figure*}[ht]
    \centering
    \setlength\abovecaptionskip{-0.2\baselineskip}
    \includegraphics[width=\textwidth]{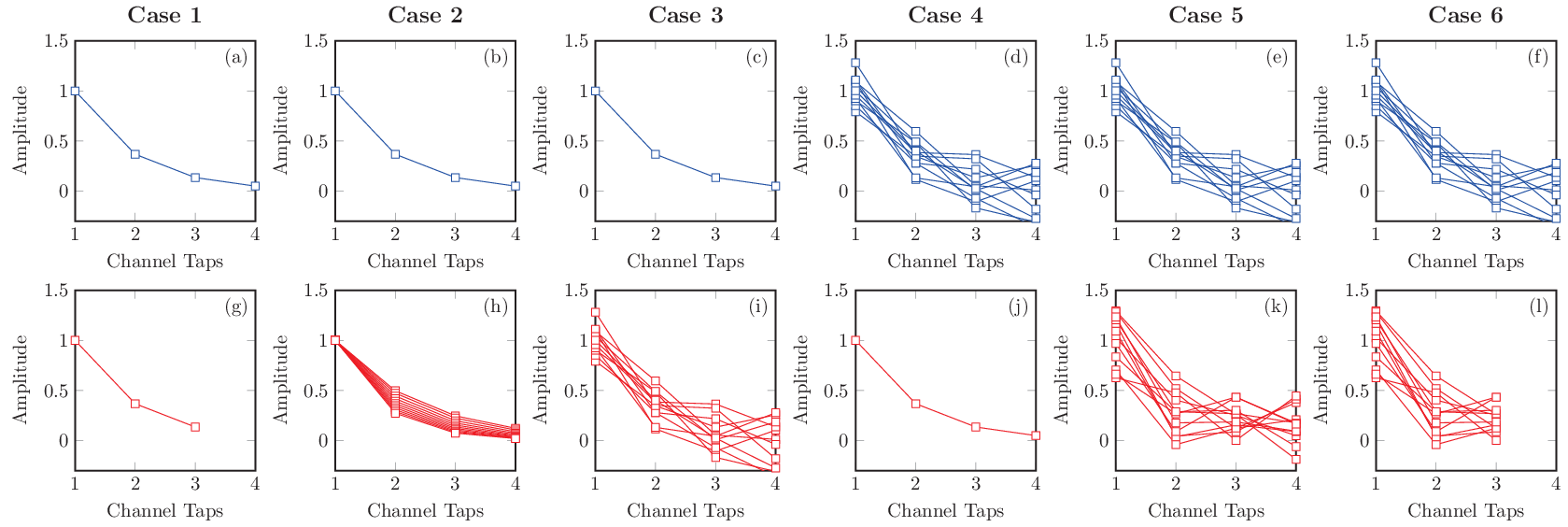} 
    \caption{Schematic diagrams (10 realizations) for the six different scenarios listed in Table~\ref{tab:exp}. The top row shows the channel taps for the simulation of the transmission channels, while the bottom row shows the estimated/training channels.}
    \label{fig:h_all}    
\end{figure*}

Our investigation aims at comprehensively evaluating the sensitivity and robustness of conventional BCJR and BCJRNet to time-varying channel conditions and/or imperfect knowledge of their parameters. As discussed in Sec.~\ref{sec:sys_model}, we consider a finite memory AWGN channel model with the input-output relationship given in \eqref{eq:channel_mod}. The input symbols $X_t$ are randomly and independently generated using binary phase shift keying ($X_t\in\{-1,1\}$, and thus, $M=2$) and sent over a channel with memory \(L\). For all numerical experiments, we fixed the signal-to-noise ratio (SNR) to 5\,dB and the channel memory to $L=4$. 

In the baseline simulation, following \cite{VitNet,Shlezinger21,Shlezinger22,Shlezinger23}, we assume that the transmission channel response obeys a time-invariant exponentially decaying profile, i.e.,
\begin{equation}\label{eq:taps}
    h_{l} = e^{-\gamma(l-1)},\qquad l={1,...,L},
\end{equation}
where $\gamma$ is the exponential decay constant. This scenario corresponds to the first three cases in Table~\ref{tab:exp}. For the time-varying channel, as shown in the bottom three rows of Table~\ref{tab:exp}, we assume that the transmission channel taps as 
\begin{equation}\label{eq:taps2}
    h_{l,t} = e^{-\gamma(l-1)} + \epsilon_{l,t}, \qquad l={1,...,L},
\end{equation}
where $\epsilon_{l,t} \in \mathcal{N}(0,\sigma_l^2)$ represents a Gaussian-distributed time-varying uncertainty with zero mean and variance $\sigma_l^2$. Moreover, we evaluate our systems with \mbox{$\gamma=\{0.5, 1, 1.5,2\}$} to represent channels with different decaying behaviors. In all simulations, the channel gain is normalized as $\bar{h}_{l} = h_l / \sqrt{\sum_p{|h_p|^2}}$.

For each case in Table~\ref{tab:exp}, a transmission channel response \eqref{eq:taps} or \eqref{eq:taps2} is employed to generate the received symbol $Y_t$. To encompass a range of imperfect CSI use-case scenarios, an estimated/training channel response is utilized to compute the likelihood function for the conventional BCJR and generate the training data for BCJRNet. In Case 1, we operate under the assumption of inaccurate channel memory knowledge $\hat{L}$. Case 2 considers the scenario involving imperfect CSI knowledge within the estimated/training decay constant $\hat{\gamma}$. In Case~3--5, we address Gaussian-distributed uncertainties within individual ISI channel taps. Lastly, Case 6 investigates the situation where the estimated/training channel taps have a Gaussian-distributed uncertainty, and the correct channel memory knowledge is also unavailable for the detectors. Figure~\ref{fig:h_all} gives an example of the transmission channel taps (top row) and the estimated/training taps (bottom row) for each of the 6 cases in Table~\ref{tab:exp}.

\begin{table}[tpb]
    \caption{Assumed channel responses in simulation}
    \renewcommand{\arraystretch}{1.3}
    \centering
    \begin{tabular}{|c|c|c|} 
        \hline
        {\bf Case}& {\bf Transmission Channel} & {\bf Estimated/Training Channel} \\
        \hline
        {\bf 1} & \multirow{3}{*}{$h_l = e^{-\gamma(l-1)}$} &  $\hat{h}_l= e^{-\gamma(l-1)}, l={1,...,\hat{L}}$ \\ \cline{1-1} \cline{3-3}
        {\bf 2} &  &  $\hat{h}_l = e^{-\hat{\gamma}(l-1)}$ \\ \cline{1-1} \cline{3-3}   
        {\bf 3} &  &  $\hat{h}_{l,t}= e^{-\gamma(l-1)} + \epsilon_{l,t}$ \\ \hline  
        {\bf 4} &   \multirow{3}{*} {$h_{l,t} = e^{-\gamma(l-1)} + \epsilon_{l,t}$} &     
        $\hat{h_l} = e^{-\gamma(l-1)}$ \\ \cline{1-1} \cline{3-3}
        {\bf 5} &  &  $\hat{h}_{l,t} = e^{-\gamma(l-1)} + \epsilon_{l,t}$ \\ \cline{1-1} \cline{3-3}
        {\bf 6} &  &  $\hat{h}_{l,t} = e^{-\gamma(l-1)} + \epsilon_{l,t}, l={1,...,\hat{L}}$ \\ \hline
    \end{tabular}
    \label{tab:exp}
\end{table}

As discussed in Sec.~\ref{sec:sys_model}, a fundamental distinction emerges in the computation of the likelihood between the conventional BCJR~\eqref{eq:p(y|s)} and BCJRNet~\eqref{eq:Bayes_rule_ML}. In the case of the conventional BCJR, the estimated channel taps $\hat{h}_{l,t}$ are used in \eqref{eq:smean} to get the estimated mean values. If the estimated coefficients $\hat{h}_{l,t}$ are different from the actual channel response $h_{l,t}$, the likelihood calculation from \eqref{eq:p(y|s)} will be sub-optimal, resulting in performance degradation. On the other hand, for BCJRNet, we use noisy channel estimates $\hat{h}_{l,t}$ to generate our training dataset $\{y_t,x_t\}_{t=1}^{T_{\text{data}}}$, and obtain the approximations \(p^{\theta}(\mathbf{s}_t|y_t)\) and \(p^{\phi}(y_t)\). Then, the likelihood is calculated based on \eqref{eq:Bayes_rule_ML}. 

For our simulations, we adopt the same NN architecture as in~\cite{Shlezinger21}, which consists of three fully connected layers ($1\times100$, $100\times50$, and $50\times M^L$), with intermediate sigmoid and ReLU activations, respectively, and a softmax output layer. The network is trained using the Adam optimizer with a learning rate 0.01. The likelihood is learned from $T_{\text{data}} = 10000$ as in~\cite{Shlezinger21}, and the presented results are averaged over 50000 Monte Carlo simulations. The simulation parameters are summarized in Table~\ref{tab:para}.

\begin{table}[tpb]
    \caption{Simulation parameters}
    \renewcommand{\arraystretch}{1.3}
    \centering
    \begin{tabular}{|c|c|} 
        \hline
        {Modulation Scheme} & {BPSK} \\
        \hline
        {SNR} & {5 dB} \\
        \hline
        {Symbol Sequence Length} & {10000} \\
        \hline
        {Monte Carlo Iterations} & {50000} \\
        \hline
        \multirow{3}{*}{NN Layers}
        & {$1\times100$ (Sigmoid)} \\
        & {$100\times50$ (ReLU)} \\
        & {$50\times 16$ (Softmax)} \\
        \hline
        {Training Epochs} & {100} \\
        \hline
        {Optimizer} & {Adam (learning rate $=0.01$)} \\
        \hline
    \end{tabular}
    \label{tab:para}
\end{table}
\section{Numerical results}\label{sec:results}

\begin{figure*}[ht]
     \centering
     \begin{subfigure}[b]{.32\textwidth}
         \centering
         \includegraphics[width=\textwidth]{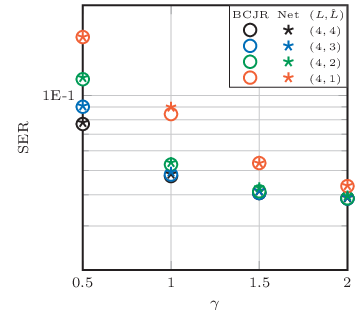} 
         \caption{Case 1}
         \label{fig:exp1} 
     \end{subfigure}
     \begin{subfigure}[b]{.32\textwidth}
         \centering
         \includegraphics[width=\textwidth]{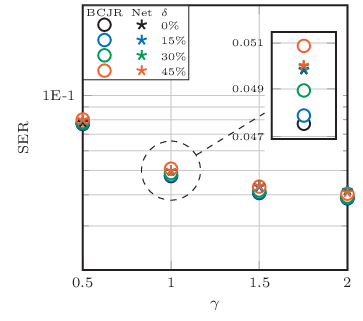} 
         \caption{Case 2}
         \label{fig:exp2} 
     \end{subfigure}
     \begin{subfigure}[b]{.32\textwidth}
         \centering
         \includegraphics[width=\textwidth]{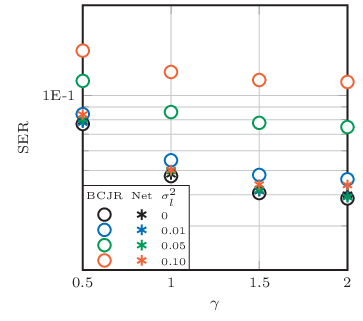} 
         \caption{Case 3}
         \label{fig:exp3}  
     \end{subfigure}
\caption{Symbol error rate (for channels operated with an ideal exponential decay profile) as a function of four different exponential decay constants $\gamma=\{0.5, 1, 1.5, 2\}$ for the conventional BCJR and BCJRNet detectors. Case 1 -- inaccurate knowledge ($\hat{L}$) of the channel memory ($L$). Case 2 -- inaccurate knowledge ($\hat{\gamma}$) of the channel parameter ($\gamma$). Case 3 -- uncertainty/deviation in the estimation/training, in terms of error variance $\sigma_l^2=\{0.01, 0.05, 0.10\}$, of the channel taps ($h_l$).}
\label{fig: 1-3}
\end{figure*}

\subsection{Time-invariant transmission channels (Cases~1--3)}
In this part, we assume that the actual channel follows an ideal exponential decay profile as in \eqref{eq:taps} and examine the detectors' performance under different imperfect CSI scenarios. For Case 1, we evaluate the receiver operation with an inaccurate channel memory assumption ($\hat{L}=1, 2, 3, 4$). As shown in Fig.~\ref{fig:exp1}, both the conventional BCJR and BCJRNet achieve their best performances when the memory of the estimated/training channel is identical to the actual channel in transmission, i.e., $(L,\hat{L})=(4,4)$. We can also observe that, in general, the error rate is lower for larger $\gamma$ since the ISI contribution is reduced. Both detectors suffer from performance degradation when they assume a shorter memory in their system design. More specifically, when $\hat{L} = 1$ the system is designed based on a memoryless channel assumption, thus resulting in a significantly increased error rate. Furthermore, when the channel is slow-decaying ($\gamma=0.5$), the impact of memory mismatch is more pronounced than in a fast-decaying ($\gamma=2$) case. This is an important observation from the system design point of view since it identifies a use-case scenario where one may assume a shorter memory at the receiver to trade off marginal performance degradation for reduced complexity. Moreover, even in the presence of inaccurate channel memory information, the data-driven solution approximates the conventional BCJR approach, which is optimal in that case.

Next, in Case 2, we consider the scenario where the knowledge of exponential decay constant $\gamma$ is inaccurate. Specifically, the goal is to simulate an operating condition that can be identified as, for example, a calibration error. For this, we draw the estimated/training $\hat{\gamma_t}$ from a uniform distribution as
\begin{equation} \label {eq: gamma}
    \hat{\gamma_t} \in \mathcal{U}[\gamma (1-\delta), \gamma (1+\delta)], 
\end{equation}
where $\delta=\{0.15,0.30,0.45\}$ can be viewed as a calibration error. Note that $\hat{\gamma_t}$ is varied for each received symbol. The symbol error rate (SER) performance is shown in Fig.~\ref{fig:exp2}. Results indicate that both detectors are robust to \(\gamma\) deviation with only small SER degradation for all considered $\gamma$ calibration errors up to $45\%$. This could be accounted for the fact that varying $\gamma$ does not cause variations in the first taps $h_{1,t}$, while the influence of the remaining taps is still exponentially decaying. Consequently, both the conventional BCJR and BCJRNet are robust to such a channel inaccuracy in terms of SER. When focusing on $\gamma=1$ (inset in Fig.~\ref{fig:exp2}), we can observe that when having perfect CSI information ($\delta=0$), the CSI-based BCJR detector is indeed the optimum solution. However, the data-driven approach exhibits smaller degradation in the presence of a slight $\gamma$ deviation. Although marginal, such an observation is important for building up an understanding of the generalization capabilities of a data-driven solution.

We proceed with our investigation to a scenario where each ISI tap of the estimated/training channel has a Gaussian-distributed uncertainty/deviation. This use-case (Case 3) is similar to the one that the authors of \cite{Shlezinger21} assumed. In addition to $\sigma_l^2=0.1$ as in \cite{Shlezinger21}, we also investigated the scenario with $\sigma_l^2=\{0.01, 0.05\}$. As mentioned in Sec.~\ref{sec:exp_design}, we normalized the channel gain, which was not done in \cite{Shlezinger21}. Therefore, there is a negative difference (e.g., $-0.55$ dB for $\gamma=1$) between the SNR assumed in this paper ($5$ dB) and the SNR in \cite{Shlezinger21}. The results of this investigation are presented in Fig.~\ref{fig:exp3}, where we indeed see that BCJRNet is able to learn from the noisy channel estimates and generalizes well to the ideal exponential power decay of the transmission channel. This introduces a significant improvement compared to the conventional BCJR. The latter fails to accurately calculate the likelihood due to a substantial mismatch between the actual and estimated Gaussian means from \eqref{eq:smean}.

\subsection{Rapidly time-varying transmission channels (Cases~4--6)}
\begin{figure*}[ht]
     \centering
     \begin{subfigure}[b]{.32\textwidth}
         \centering
         \includegraphics[width=\textwidth]{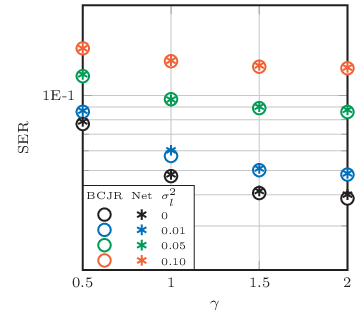} 
         \caption{Case 4}
         \label{fig:exp4} 
     \end{subfigure}
     \begin{subfigure}[b]{.32\textwidth}
         \centering
         \includegraphics[width=\textwidth]{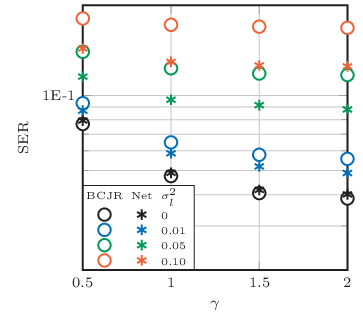} 
         \caption{Case 5}
         \label{fig:exp5} 
     \end{subfigure}
     \begin{subfigure}[b]{.32\textwidth}
         \centering
         \includegraphics[width=\textwidth]{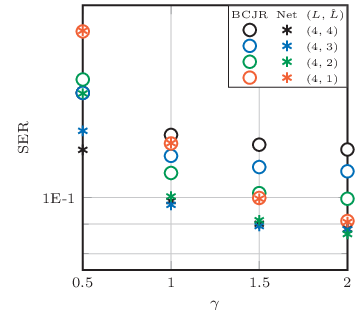} 
         \caption{Case 6}
         \label{fig:exp6}  
     \end{subfigure}
\caption{Symbol error rate (for channels operated with rapidly time-varying channel taps with variances $\sigma_l^2=\{0.01, 0.05, 0.10\}$ for Case 4 and 5, and $\sigma_l^2=\{0.05\}$ for Case 6) as a function of four different exponential decay constants $\gamma=\{0.5, 1, 1.5, 2\}$ for the conventional BCJR and BCJRNet detectors. Case 4 -- ideal exponential decay estimated/training channel. Case 5 -- rapidly time-varying estimated/training channel (with identical variances as in the transmission channel). Case 6 -- rapidly time-varying estimated/training channel ($\sigma_l^2=0.05$) as well as inaccurate knowledge ($\hat{L}$) of the channel memory ($L$).}
\label{fig:4-6}
\end{figure*}

To further extend the robustness investigation, we focus on cases where the transmission channel is rapidly time-varying in Cases 4 -- 6. This scenario is more commonly encountered in practical wireless communication systems. Figure~\ref{fig:exp4} shows the results of Case 4, where the estimated/training channel follows an ideal exponential decay profile (the average profile of the actual channel). We can observe that BCJRNet, similar to the conventional BCJR, fails to generalize for operation in rapidly time-varying conditions.

Figure~\ref{fig:exp5} shows the SER result when both the actual and estimated/training channels are rapidly varying in time (Case 5). Although both detectors perform relatively poorly in these conditions, BCJRNet achieves a lower error rate. It is interesting to observe that, comparing the results from Figs.~\ref{fig:exp4} and~\ref{fig:exp5}, it can be concluded that BCJRNet converges to identical solutions in both situations, and thus, performance does not differ. 

Lastly, in Case 6, we introduce imperfect CSI in channel taps and memory. The result, shown in Fig.~\ref{fig:exp6} indicates that for the conventional BCJR, the impact of inaccurate memory assumption is greater when the channel is slow-decaying ($\gamma=0.5$). Interestingly, in the fast-decaying case, the uncertainty/deviation of channel impulse response dominates the system performance. Hence, the imperfect knowledge of the channel memory has a mitigating effect on the degradation from mismatched taps. On the other hand, BCJRNet is relatively robust, especially for a fast-decaying channel.

\section{Conclusions}\label{sec:conc}

Our study highlighted the importance of considering the robustness of data-driven communication systems to varying channel conditions and inaccurate channel parameter knowledge between learning and operation stages. It aimed to identify suitable application scenarios for such receivers. In particular, we investigated BCJRNet comprehensively and compared it to the conventional BCJR algorithm. Our findings reveal that, unlike the CSI-based implementation, the data-driven approach manages to generalize for inaccurate knowledge of channel taps for a stationary transmission channel. However, the performance of BCJRNet experiences a pronounced deterioration when the transmission channel is also rapidly time-varying. Furthermore, the assumption of memory emerges as an important factor for both conventional and data-driven BCJR algorithms, particularly in cases where the channel is slowly decaying. Also, our observations indicate that, in the case of the conventional BCJR detector, underestimating the channel memory, albeit mismatched with the actual channel, has the potential to mitigate the detrimental impact of uncertainties in channel tap coefficients. Conversely, BCJRNet consistently improves performance when the memory assumption approaches the operating channel.

\section*{Acknowledgment}
This work is carried out under the RAISE project which is a collaboration between NXP Semiconductors and Eindhoven University of Technology.

\bibliographystyle{IEEEtran}
\bibliography{ref}

\vspace{12pt}
\end{document}